\documentclass[aps,prf,12pt]{revtex4-1}

\usepackage[T1]{fontenc}		
\usepackage[utf8]{inputenc}		
\usepackage{lmodern}			
\usepackage[english]{babel}
\usepackage{etoolbox}

\usepackage{booktabs}	
\usepackage{enumitem}

\usepackage[usenames,dvipsnames]{xcolor} 

\usepackage[colorlinks]{hyperref}

\usepackage{scrextend}

\usepackage{graphicx}
\usepackage{tikz,pgfplots}		
\usetikzlibrary{arrows,fit,shapes}

\newlength\figureheight
\newlength\figurewidth

\usepackage{subfig}				

\usepackage{amsmath}

\usepackage{amssymb}
\usepackage{mathtools}
\usepackage[squaren]{SIunits}	

\usepackage{caption}

\usepackage{cases}

\usepackage{newtxtext,newtxmath}

\usepackage{bm}
\usepackage{ulem}

\newcommand{\st}[2][]{_{\text{#2}{#1}}} 
\newcommand{\spt}[1]{^{\text{#1}}}

\newcommand{\bvec}[1]{{\bm{#1}}}

\newif\ifshowchanges
\showchangesfalse

\ifshowchanges
\newcommand{\myHighlight}[1] {\textcolor{ForestGreen}{#1}}
\newcommand{\myDelete}[1] {\textcolor{red}{\sout{#1}}}  
\newcommand{\myOpDelete}[1] {#1}

\else
\newcommand{\myHighlight}[1] {#1}
\newcommand{\myDelete}[1] {}
\newcommand{\myOpDelete}[1] {}

\fi
\graphicspath{{./figures/}}

\begin{document}


\title{Data-assisted, physics-informed propagators for recurrent flows}
\author{T. Lichtenegger}
\email{thomas.lichtenegger@jku.at}
\affiliation{Department of Particulate Flow Modelling, Johannes Kepler University Linz, Altenberger Str.\ 69, 4040 Linz, Austria}



\date{\today}

\begin{abstract}
A novel approach to simulate dynamic, recurrent flows with very large time steps is presented. Data-driven forecasts based on the method of analogues are made employing a set of precomputed time series segments of flow fields. These predictions are then corrected with the evolved deviation between initial flow state and its nearest neighbor in the database. The evolution of these deviations is obtained from the convolution with the propagator of the linearized Navier-Stokes equations, where linearization takes place around the most similar state in the database. The error $E\st{lin}$ due to this approximation implies the appropriate distance function to determine nearest-neighbor fields.
We tested the method on the simple case of 2D vortex shedding behind a cylinder at Reynolds numbers in the range of 200 with step sizes of up to 200 times larger than those used to generate the underlying data with the PISO algorithm. Short-term forecasts showed a high degree of accuracy with errors correlating well with $E\st{lin}$ for a number of case variations. But also after many steps, the predicted flow fields remained in close agreement with high-fidelity CFD calculations with relative errors on the order of $1\%$ for the temporal averages. Furthermore, excellent conservation behavior of global properties like total momentum and kinetic energy was observed.
\end{abstract}

\keywords{data-assisted simulations; physics-informed ansatz; propagator approach; recurrent flows}

\maketitle

\section{Introduction}
\label{S:Intro}
Transient fluid mechanical problems and their numerical treatment are connected to small time steps with expensive calculations in each of them. Even though hardware resources have increased significantly to allow for massively parallel computations, it is still very demanding to simulate highly dynamic flows over long durations. For this reason, temporal multi-scale problems involving very fast and slow processes have remained a serious challenge.

Over the last few years, data-driven numerical techniques involving elements of machine learning (ML) have started to enter fluid mechanics. An overview of the vast number of methods and their applications can be found in the review of Brunton~et~al.~\cite{brunton2019}. Various strategies, many of them built upon deep neural networks (NNs), to learn certain aspects of flow dynamics from data were devised. Generative adversarial networks (GANs) were trained to make increasingly realistic predictions about the evolution~\cite{Cheng2020,Akkari2020,Quilodran2023} and about sub-resolution structures~\cite{Deng2019,Janssens2020,Yousif2022} of flows. Others extracted characteristic patterns and reduced the dimensionality of the problem to derive reduced-order models (ROMs). Besides the classical combination of proper orthogonal decomposition (POD) with a Galerkin approach~\cite{Bergmann2009,Lorenzi2016,Stabile2017,Cavalieri2022}, more recent works employed autoencoders~\cite{agostini2020,Fukami2020,murata2020,Fukami2021,Page2021,DeJesus2023} to allow for a non-linear decomposition. Equations of motion (EOMs) acting in these reduced coordinates~\cite{Brunton2016,Champion2019,Loiseau2020,Guan2021} come with significantly lower computational costs than those of the original problem.

The impact of combining data with the underlying laws of physics was demonstrated by Raissi~et~al.~\cite{raissi2018c,raissi2019,raissi2020} in terms of physics-informed neural networks (PINNs), which have proven to be extremely powerful tools for ill-posed and inverse problems and have been applied to a wide range of topics~\cite{Raissi2019b,Yang2019,Sahli2020,Cai2021,Reyes2021}.
Including the governing EOMs into the loss function of a NN allowed to make accurate predictions with comparatively little data as opposed to techniques which are not aware of any physical constraints. An overview of the rapidly growing field of physics-informed ML is provided by the review of Karniadakis~et~al.~\cite{Karniadakis2021}.

In stark contrast to such elaborate techniques, we developed the deliberately simple yet very fast method recurrence CFD (rCFD) specifically targeted at temporal multi-scale problems in recurrent flows like fluidized beds~\cite{Lichtenegger2017b} or turbulent jets~\cite{Abbasi2020b}. Using a database of prerecorded field time series, rCFD performs large-step predictions by iterating the method of analogues~\cite{cecconi2012}. This way, flow sequences of almost arbitrary length can be easily generated, which can serve as a dynamic background for slow long-term processes to take place. Especially for passive transport, rCFD enabled extremely fast simulations several orders of magnitude~\cite{Pirker2018} faster than the underlying CFD calculations \myHighlight{after the initial construction of the database with said methods. Of course, rCFD does not reach the same level of accuracy as classic CFD methods but can provide results in semi-quantitative agreement with them.
Conversely, standard CFD algorithms can be accelerated to some degree by using larger steps and decreasing the required accuracy, but could hardly achieve similar CPU times because they would still have to solve the coupled velocity and pressure equations. However, it cannot be denied that open, challenging questions for rCFD persist such as the handling of high Reynolds number flows, where the velocity field varies over time scales beyond the scope of a single database~\cite{Lichtenegger2022}}.

For the case of long-term simulations, it is particularly important but also challenging to ensure that the governing laws of physics are still satisfied after long durations. While physics-informed ML models include the EOMs directly in their training, rCFD uses only flow fields from its physically valid database. Therefore, rCFD is more or less bound to the conditions at which the database was created, but at the same time does not run into any danger of leaving the underlying attractor even after a huge number of steps. Physics-informed ML techniques on the other hand can offer more flexibility but might struggle with multi-scale problems~\cite{Karniadakis2021,wang2022}.

In this work, we propose a novel approach that combines the benefits of flexibility and stability in a clearly interpretable fashion.
Large-step predictions are made based on the method of analogues. If the exactly same state as that for which the forecast is to be computed can found in the database, one could infer the evolution without any uncertainty. However, in general there will be deviations between the state under consideration and its nearest neighbor from the database, e.g., due to different boundary conditions or changes of internal parameters. Clearly, these differences need to be taken into account to correct the predictions.
To this end, we derive equations for the propagation of the deviations by linearizing the full, non-linear Navier-Stokes equations around the nearest neighbor state.
\myHighlight{In contrast to (partly) implicit techniques like backward differentiation formulas (BDF) or explicit singly diagonal implicit Runge-Kutta (ESDIRK) methods~\cite{Bijl2002,Forti2015,Loy2019}, which can tolerate large time steps, too, the present algorithm is formulated in an almost completely explicit fashion (cf.~Sec.~\ref{S:performance}), which gives it the potential for much faster simulations.}

Related ideas were put forward by Cacuci~et~al.~\cite{Cacuci1988,Cacuci1989}, who developed a Green's function formalism for non-linear problems. However, they obtained self-consistent integral equations whereas we restrict ourselves to a first-order expansion that leads to a simple, explicit scheme in terms of a deviation propagator. Gin~et~al.~\cite{Gin2021} used autoencoders to transform  boundary value problems into linear coordinates where a Green's function can be found, but a physics-informed extension to time-dependent settings has not been attempted yet.
Boull{\'e}~et~al.~\cite{Boulle2022} used NNs to learn the response of a system excited with a Gaussian process. Their approach may also be applied to non-linear EOMs to obtain the Green's function for an approximate linear model. It is an open question to which extent this simplification can account for a non-linearity in transient problems over long durations. \myHighlight{Similarly, physics-informed deep operator networks~\cite{Lu2021,li2021b,Wang2021b,Wang2023} advance a flow state in time by learning the underlying, non-linear EOMs, which becomes increasingly challenging for large, multi-scale systems.}
In contrast, our method allows for a clean separation into a non-linear sub-problem and a linear one for which a propagator can be computed, but relies on a database of similar flow fields. Therefore, it is applicable when similar states are repeatedly encountered. This comprises transio-recurrent dynamics and cases where the effects of parameter variation, e.g., for the purpose of optimization, are of interest. We remark that the combination of nearest-neighbor prediction with deviation propagation may also be applied to other non-linear problems than the Navier-Stokes equations. 

The paper is organized as follows: The theoretical basis of the deviation-propagation formalism is explained in Sec.~\ref{S:Theory}. In Sec.~\ref{S:Setup}, we describe the test case that serves to illustrate our method with results presented in Sec.~\ref{S:Results}. Finally, we draw conclusions and discuss limitations in Sec.~\ref{S:Conclusion}.

\section{Theoretical background}
\label{S:Theory}
\subsection{Linearized propagator ansatz}
\myHighlight{We investigate the dynamics of an incompressible fluid of viscosity $\nu$ with velocity $\bvec{u}(\bvec{r},t)$ and pressure $p(\bvec{r},t)$ in a domain $\Omega$ governed by the Navier-Stokes equations
\begin{align}
&\nabla\cdot\bvec{u} = 0 \label{eq:ns1} \\
&\frac{\partial}{\partial t} \bvec{u} + \nabla\cdot \bvec{u}\bvec{u} = -\nabla p + \nu\nabla^2 \bvec{u}.\label{eq:ns2}
\end{align}
}
Given two sufficiently similar flow fields with sufficiently similar values on the boundaries, which are subject to the same EOMs, it is reasonable to assume that they will evolve in a similar fashion for some (short) time period. More specifically, assuming that the evolution of a reference state $\bvec{u}\st{ref}(\bvec{r},0<t'\leq t),\, p\st{ref}(\bvec{r},0<t'\leq t)$ is known up to time $t$ and available in a database of previously calculated time series segments, we want to model the dynamics observed for similar initial conditions
\begin{align}
&\bvec{u}(\bvec{r},0) = \bvec{u}\st{ref}(\bvec{r},0) + \delta \bvec{u}(\bvec{r},0) \\
&p(\bvec{r},0) = p\st{ref}(\bvec{r},0) + \delta p(\bvec{r},0)
\end{align}
with similar boundary values. For the sake of simplicity, we assume that the boundary \myHighlight{of $\Omega$,} $\partial\Omega = \partial\Omega_u \cup \partial\Omega_p$\myHighlight{,} consists of patches $\partial\Omega_u$ with fixed value of velocity and gradient of pressure and of patches $\partial\Omega_p$ with fixed value of pressure and gradient of velocity so that
\begin{equation}
\begin{rcases}
\bvec{u}(\bvec{r}_b,t) = \bvec{u}\st{ref}(\bvec{r}_b,t) + \delta \bvec{u}(\bvec{r}_b,t) \\
\bvec{n}\cdot \nabla p(\bvec{r}_b,t) = \bvec{n}\cdot \nabla p\st{ref}(\bvec{r}_b,t) + \delta p'(\bvec{r}_b,t)
\end{rcases} \,\, \text{for}\,\, \bvec{r}_b\in \partial\Omega_u
\label{eq:BC0}
\end{equation}
and
\begin{equation}
\begin{rcases}
\bvec{n}\cdot \nabla\bvec{u}(\bvec{r}_b,t) = \bvec{n}\cdot \nabla\bvec{u}\st{ref}(\bvec{r}_b,t) + \delta \bvec{u}'(\bvec{r}_b,t) \\
 p(\bvec{r}_b,t) = p\st{ref}(\bvec{r}_b,t) + \delta p(\bvec{r}_b,t)
\end{rcases}\,\, \text{for}\,\, \bvec{r}_b\in \partial\Omega_p.
\label{eq:BC1}
\end{equation}
As long as all deviations indicated by a leading $\delta$ are small, we can make a linear ansatz
\begin{align}
\bvec{u}(\bvec{r},t) = &\bvec{u}\st{ref}(\bvec{r},t) \nonumber\\
&+\int_{\Omega} d^3r' \bvec{K}_{\bvec{u}\bvec{u}}(\bvec{r},\bvec{r}',t)\cdot \delta\bvec{u}(\bvec{r}',0)\nonumber\\
&+\int_{\Omega} d^3r' \bvec{K}_{\bvec{u}p}(\bvec{r},\bvec{r}',t) \delta p(\bvec{r}',0)\label{eq:uansatz}\\
p(\bvec{r},t) = &p\st{ref}(\bvec{r},t) \nonumber\\
&+\int_{\Omega} d^3r' \bvec{K}_{p\bvec{u}}(\bvec{r},\bvec{r}',t)\cdot \delta\bvec{u}(\bvec{r}',0)\nonumber\\
&+\int_{\Omega} d^3r' \bvec{K}_{pp}(\bvec{r},\bvec{r}',t) \delta p(\bvec{r}',0),
\label{eq:pansatz}
\end{align}
where integration is carried out over the whole domain $\Omega$ including the boundaries. The latter part is essential because of delta functions located on them (cf.\ the following sections).

While the first term on the right-hand sides of Eqs.~\eqref{eq:uansatz} and \eqref{eq:pansatz} represents the evolution of the reference state, the others account for the effect of the deviations of velocity and pressure. The kernels $\bvec{K}_{\bvec{u}\bvec{u}}$, $\bvec{K}_{\bvec{u}p}$, $\bvec{K}_{p\bvec{u}}$, and $\bvec{K}_{pp}$ can be regarded as yet to be determined components of the\myDelete{deviation} propagator that evolves a small deviation $\delta\bvec{u}(0),\,\delta p(0)$ from a reference state over a duration $t$ into $\delta\bvec{u}(t),\,\delta p(t)$. They are closely related to the Green's function of the system (cf.~Sec.~\ref{S:eoms}).

\myHighlight{As long as the database is sufficiently extensive, a similar reference state for $\bvec{u}(t),\, p(t)$ computed by Eqs.~\eqref{eq:uansatz} and \eqref{eq:pansatz} can be found and the procedure repeated to obtain $\bvec{u}(2t),\, p(2t)$ and so on.}

\subsection{Initial and boundary conditions of the propagators}
The initial conditions
\begin{align}
&\bvec{K}_{\bvec{u}\bvec{u}}(\bvec{r},\bvec{r}',0) = \bvec{1}\delta(\bvec{r}-\bvec{r}')\label{eq:Xuu_init}\\
&\bvec{K}_{\bvec{u}p}(\bvec{r},\bvec{r}',0) = \bvec{0}\label{eq:Xup_init}
\end{align}
follow from the requirement that $\bvec{u}(\bvec{r},t \rightarrow 0) = \bvec{u}\st{ref}(\bvec{r},t\rightarrow 0) +  \delta\bvec{u}(\bvec{r},0)$ in Eq.~\eqref{eq:uansatz}. No initial conditions are needed for $\bvec{K}_{p\bvec{u}}$ and $\bvec{K}_{pp}$ because no time derivatives of pressure are computed in incompressible flow.

In addition to the initial conditions, one has to define appropriate boundary conditions. They need to be specified only for the first coordinate of the propagators because no spatial derivatives with respect to the second coordinates are calculated (cf.~Sec.~\ref{S:eoms}). 
If $\bvec{r}$ approaches a point $\bvec{r}_b$ on the boundary, one finds that
\begin{align}
\begin{rcases}
&\bvec{K}_{\bvec{u}\bvec{u}}(\bvec{r}\rightarrow \bvec{r}_b,\bvec{r}') = \bvec{1}\delta(\bvec{r}_b-\bvec{r}')\\
&\bvec{K}_{\bvec{u}p}(\bvec{r}\rightarrow \bvec{r}_b,\bvec{r}') = \bvec{0},
\end{rcases} \,\, \text{for}\,\, \bvec{r}_b\in \partial\Omega_u \label{eq:BCDirichletU}
\end{align}
and
\begin{equation}
\begin{rcases}
&\bvec{n}\cdot\nabla\bvec{K}_{\bvec{u}\bvec{u}}(\bvec{r}\rightarrow \bvec{r}_b,\bvec{r}') = \bvec{n}\cdot\nabla \bvec{1}\delta(\bvec{r}_b-\bvec{r}')\label{eq:BCNeumannU}\\
&\bvec{n}\cdot\nabla \bvec{K}_{\bvec{u}p}(\bvec{r}\rightarrow \bvec{r}_b,\bvec{r}') = \bvec{0}
\end{rcases} \,\, \text{for}\,\, \bvec{r}_b\in \partial\Omega_p.
\end{equation}
have to hold such that the velocity field according to ansatz Eq.~\eqref{eq:uansatz} satisfies the boundary conditions Eqs.~\eqref{eq:BC0} and \eqref{eq:BC1}. Similarly, if the pressure components of the propagator obey
\begin{align}
\begin{rcases}
& \bvec{n}\cdot\nabla\bvec{K}_{p\bvec{u}}(\bvec{r}\rightarrow \bvec{r}_b,\bvec{r}') = \bvec{0}\\
& \bvec{n}\cdot\nabla\bvec{K}_{pp}(\bvec{r}\rightarrow \bvec{r}_b,\bvec{r}') = \bvec{n}\cdot\nabla \delta(\bvec{r}_b-\bvec{r}')
\end{rcases} \,\, \text{for}\,\, \bvec{r}_b\in \partial\Omega_u\label{eq:BCNeumannP}
\end{align}
and
\begin{align}
\begin{rcases}
& \bvec{K}_{p\bvec{u}}(\bvec{r}\rightarrow \bvec{r}_b,\bvec{r}') = \bvec{0}\\
& \bvec{K}_{pp}(\bvec{r}\rightarrow \bvec{r}_b,\bvec{r}') = \delta(\bvec{r}_b-\bvec{r}')
\end{rcases} \,\, \text{for}\,\, \bvec{r}_b\in \partial\Omega_p,\label{eq:BCDiricletP}
\end{align}
then the predicted pressure field agrees with the prescribed boundary conditions.

\subsection{Propagator EOMs}
\label{S:eoms}
For consistency reasons, the linear ansatz Eqs.~\eqref{eq:uansatz} and \eqref{eq:pansatz} implies that the propagators are determined from the linearized Navier-Stokes equations, where linearization \myHighlight{of Eqs.~\eqref{eq:ns1} and \eqref{eq:ns2}} takes place around $\bvec{u}\st{ref}(\bvec{r},t), p\st{ref}(\bvec{r},t)$,
\begin{align}
&\nabla\cdot \delta \bvec{u} = 0 \label{eq:linns1}\\
& \frac{\partial}{\partial t}\delta \bvec{u} + \nabla\cdot \delta \bvec{u}\bvec{u}\st{ref} + \nabla\cdot \bvec{u}\st{ref}\delta \bvec{u} = -\nabla \delta p + \nu\nabla^2\delta \bvec{u}.\label{eq:linns2}
\end{align}
This way, we have moved all the complexity due to the non-linearity of the full Navier-Stokes equations into the creation of a representative database of flow field time series segments. Any deviation from these prerecorded states is dealt with by the propagators in a linearized fashion.

In the above approximation, one neglects the second-order term $\nabla\cdot \delta \bvec{u}\delta \bvec{u}$.
\myHighlight{This is valid as long as its magnitude satisfies
\begin{equation}
\big|\nabla\cdot \delta \bvec{u}\delta \bvec{u}\big| \ll |\nabla\cdot \delta \bvec{u}\bvec{u}\st{ref} + \nabla\cdot \bvec{u}\st{ref}\delta \bvec{u}|,\label{eq:lincriterion}
\end{equation}
which makes it}
\myDelete{Because its magnitude determines the validity of the linearized approach, it is}a reasonable indicator to identify a sufficiently similar reference state\myDelete{, e.g.,} in terms of
\begin{equation}
E\st{lin} \equiv \int_{\Omega} d^3r \big|\nabla\cdot \delta \bvec{u}\delta \bvec{u}\big|.\label{eq:errindicator}
\end{equation}
\myHighlight{Furthermore, Eq.~\eqref{eq:lincriterion} allows to estimate an upper bound for the spacing of reference states. If $\bvec{u}(\bvec{r},t_n)$ is the $n$-th reference velocity field, the next one needs to be added to the database before $\delta \bvec{u} = \bvec{u}(\bvec{r},t_{n+1})-\bvec{u}(\bvec{r},t_n)$ and $\bvec{u}\st{ref} = \bvec{u}(\bvec{r},t_n)$ violate Eq.~\eqref{eq:lincriterion}. Of course, this criterion is only sufficient if the large-step simulations are carried out at the conditions of the database. For different boundary conditions or internal parameter values, $\delta \bvec{u}$ cannot be made arbitrarily small by refining the distance between reference states. Instead, its magnitude is limited from below by the distance between the current conditions and those of the database. For such cases, Eq.~\eqref{eq:lincriterion} may be repeatedly checked during large-step simulations to assess the validity of the propagator approach. 
}

If we now insert the ansatz functions and keep in mind that $\delta\bvec{u}(\bvec{r},0)$ and $\delta p(\bvec{r},0)$ are independent from each other, we obtain separate equations for the propagation of velocity variations $\bvec{K}_{\bvec{u}\bvec{u}}$, $\bvec{K}_{p\bvec{u}}$ and for the propagation of pressure variations $\bvec{K}_{\bvec{u}p}$, $\bvec{K}_{pp}$.

First, we deal with the former case. In order for Eq.~\eqref{eq:uansatz} to satisfy the continuity equation, one needs
\begin{equation}
 \int_{\Omega} d^3r' \nabla\cdot \bvec{K}_{\bvec{u}\bvec{u}}(\bvec{r},\bvec{r}',t)\cdot \delta\bvec{u}(\bvec{r}',0) = 0.\label{eq:divfreeop}
\end{equation}
Although it would be tempting to assume $\nabla\cdot \bvec{K}_{\bvec{u}\bvec{u}}(\bvec{r},\bvec{r}',t)=0$, this equation is clearly invalid 
as it conflicts with the initial condition Eq.~\eqref{eq:Xuu_init}.
The reason for this seeming contradiction lies in the fact that $\bvec{K}_{\bvec{u}\bvec{u}}(\bvec{r},\bvec{r}',t)$ is defined to only operate on physically valid, solenoidal states $\delta\bvec{u}(\bvec{r},0)$. Hence, one can conclude from Eq.~\eqref{eq:divfreeop} only that ${\nabla\cdot \bvec{K}_{\bvec{u}\bvec{u}}(\bvec{r},\bvec{r}',t)= \bvec{\mathcal N}(\bvec{r},\bvec{r}',t)}$ with some operator $\bvec{\mathcal N}$ the evaluates to 0 if applied to a divergence-free flow field.

\myHighlight{In this context, it might be advantageous to expand $\bvec{K}_{\bvec{u}\bvec{u}}(\bvec{r},\bvec{r}',t)$ in a basis of divergence-free finite elements~\cite{Gustafson1983,Ye1997,Schroeder2017,Allendes2021} with respect to $\bvec{r}'$ where the $\bvec{r}$-dependent coefficients would be indeed solenoidal. Upon representing the velocity field in the same basis, large-step predictions would be divergence-free per construction. However, in the present investigation we restrict ourselves to the simpler finite-volume discretization, for which}
\myDelete{Because of this distinction,} we first derive the pressure equation from Eqs.~\eqref{eq:linns1}
 and ~\eqref{eq:linns2} and then eliminate any terms of the form $\nabla\cdot\delta \bvec{u}(\bvec{r},t)$ (which are identical 0) from it and from the momentum equation. Next, we insert the ansatz functions Eqs.~\eqref{eq:uansatz} and ~\eqref{eq:pansatz} and postulate that the resulting equations hold for every $\bvec{r}'$ separately. For the momentum equation, one obtains
\begin{align}
& \frac{\partial}{\partial t} \bvec{K}_{\bvec{u}\bvec{u}}
+ \bvec{K}_{\bvec{u}\bvec{u}}\cdot\nabla \bvec{u}\st{ref} + \bvec{u}\st{ref}\cdot\nabla \bvec{K}_{\bvec{u}\bvec{u}} = -\nabla \bvec{K}_{p\bvec{u}} + \nu\nabla^2\bvec{K}_{\bvec{u}\bvec{u}}\label{eq:momU}
\end{align}
or written component-wise 
\begin{align}
& \frac{\partial}{\partial t} K^{(\bvec{u}\bvec{u})}_{ik}
+ K^{(\bvec{u}\bvec{u})}_{jk}u\spt{(ref)}_{i,j}
+ u\spt{(ref)}_{j} K^{(\bvec{u}\bvec{u})}_{ik,j} =
- K^{(p\bvec{u})}_{k,i}
+\nu K^{(\bvec{u}\bvec{u})}_{ik,jj},
\end{align}
where indices after a comma represent derivatives.
It is important to note that the derivatives in Eq.~\eqref{eq:momU} only act on the first coordinate of the propagators and that they are decoupled with regard to the second index. Hence, to obtain the tensorial $\bvec{K}_{\bvec{u}\bvec{u}}$ that depends on two spatial coordinates, one may solve a large number of vectorial equations in 3D.

The close connection of the propagators with the Green's functions
\begin{align}
&\bvec{G}_{\bvec{u}\bvec{u}}(\bvec{r},\bvec{r}',t) \equiv \theta(t)\bvec{K}_{\bvec{u}\bvec{u}}(\bvec{r},\bvec{r}',t) \\
&\bvec{G}_{p\bvec{u}}(\bvec{r},\bvec{r}',t) \equiv \theta(t)\bvec{K}_{p\bvec{u}}(\bvec{r},\bvec{r}',t),
\end{align}
where $\theta(t)$ is the Heaviside function, can be easily confirmed on Eq.~\eqref{eq:momU} using the initial condition Eq.~\eqref{eq:Xuu_init}. One finds
\begin{equation}
\frac{\partial}{\partial t} \bvec{G}_{\bvec{u}\bvec{u}}
+ \bvec{G}_{\bvec{u}\bvec{u}}\cdot\nabla \bvec{u}\st{ref} + \bvec{u}\st{ref}\cdot\nabla \bvec{G}_{\bvec{u}\bvec{u}} +\nabla \bvec{G}_{p\bvec{u}} - \nu\nabla^2\bvec{G}_{\bvec{u}\bvec{u}} = \delta(t)\delta(\bvec{r}-\bvec{r}')\bvec{1},
\end{equation}
which is the very definition of a Green's function.

Having derived the \myDelete{propatator}\myHighlight{propagator} momentum equation, that for pressure takes the form
\begin{equation}
\text{Tr}\Big[\big(\nabla\bvec{K}_{\bvec{u}\bvec{u}} \big) \cdot \big(\nabla \bvec{u}\st{ref}\big)\Big] + \text{Tr}\Big[\big(\nabla \bvec{u}\st{ref}\big) \cdot \big(\nabla\bvec{K}_{\bvec{u}\bvec{u}}\big)\Big] = - \nabla^2 \bvec{K}_{p\bvec{u}} \label{eq:pU}
\end{equation}
and
\begin{align}
K^{(\bvec{u}\bvec{u})}_{jk,i}u\spt{(ref)}_{i,j}
+
K^{(\bvec{u}\bvec{u})}_{ik,j}u\spt{(ref)}_{j,i}
=
- K^{(p\bvec{u})}_{k,ii},
\end{align}
respectively. Ideally, a propagator obeying this equation leads to solenoidal velocity fields. However, numerical inaccuracies that accumulate over the course of many steps, e.g., because of discretization errors, could occur and might need to be corrected by solving a Poisson equation for the neglected contribution to pressure that restores solenoidality. As long as the effect remains small, convergence will be fast and the impact on the numerical costs not too severe.

The time evolution due to pressure variations is governed by similar equations with minor deviations.
Although the continuity equation 
\begin{equation}
\nabla \cdot \bvec{K}_{\bvec{u}p} = 0,
\end{equation}
may be written down explicitly because the pressure field $\delta p(\bvec{r}',0)$ is not constrained as the velocity, the momentum equation
\begin{align}
&\frac{\partial}{\partial t} \bvec{K}_{\bvec{u}p}
+ \bvec{K}_{\bvec{u}p}\cdot\nabla \bvec{u}\st{ref} + \bvec{u}\st{ref}\cdot\nabla \bvec{K}_{\bvec{u}p} = -\nabla \bvec{K}_{pp} + \nu\nabla^2\bvec{K}_{\bvec{u}p} \label{eq:momp}\\
& \frac{\partial}{\partial t} K^{(\bvec{u}p)}_{i}
+ K^{(\bvec{u}p)}_{j}u\spt{(ref)}_{i,j}
+ u\spt{(ref)}_{j} K^{(\bvec{u}p)}_{i,j} =
- K^{(pp)}_{,i}
+\nu K^{(\bvec{u}p)}_{i,jj}. 
\end{align}
and that for pressure
\begin{align}
& \text{Tr}\Big[\big(\nabla\bvec{K}_{\bvec{u}p}\big)\cdot\big( \nabla \bvec{u}\st{ref}\big)\Big] + \text{Tr}\Big[\big(\nabla \bvec{u}\st{ref}\big) \cdot \big(\nabla\bvec{K}_{\bvec{u}p}\big)\Big] = - \nabla^2 \bvec{K}_{pp}\label{eq:pp}\\
& K^{(\bvec{u}p)}_{j,i}u\spt{(ref)}_{i,j}
+
K^{(\bvec{u}p)}_{i,j}u\spt{(ref)}_{j,i}
=
- K^{(pp)}_{,ii},
\end{align}
are of identical structure as for the evolution due to velocity variations.

Unless $\bvec{r}' \rightarrow \bvec{r}_b \in \partial\Omega_{u,p}$, these equations are solved by $\bvec{K}_{\bvec{u}p}(\bvec{r},\bvec{r}',t) = \bvec{0}$ and ${\bvec{K}_{pp}(\bvec{r},\bvec{r}',t) = 0}$, which is a mere consequence from the fact that pressure inside the domain follows from the velocity field and can only be specified independently on the boundary. 

If the pressure on $\partial \Omega_p$ is the same as that of the reference solution, it is obvious that no contribution to the velocity and pressure prediction arises from $\partial \Omega_p$.
Furthermore, for a pressure field that has the same gradient on $\partial\Omega_u$ as the reference state and hence a vanishing gradient of the deviation, these contributions of the propagator cancel each other out. This is a consequence from the following fact: If $\bvec{K}_{\bvec{u}p}(\bvec{r},\bvec{r}_b \in \partial\Omega_u,t)$, $\bvec{K}_{pp}(\bvec{r},\bvec{r}_b \in \partial\Omega_u,t)$ solve the EOMs, so do $-\bvec{K}_{\bvec{u}p}(\bvec{r},\bvec{r}_b + \epsilon\bvec{n},t)$, $-\bvec{K}_{pp}(\bvec{r},\bvec{r}_b + \epsilon\bvec{n},t)$ for $\epsilon \rightarrow 0$ with an inwards pointing surface normal unit vector $\bvec{n}$. It is straight forward to see that Eqs.~\eqref{eq:momp} and \eqref{eq:pp} remain invariant under change of sign, and that the initial condition Eq.~\eqref{eq:Xup_init} and boundary conditions Eqs.~\eqref{eq:BCDirichletU} and \eqref{eq:BCNeumannU} for $\bvec{K}_{\bvec{u}p}$ hold. While the boundary condition Eq.~\eqref{eq:BCDiricletP} on $\partial\Omega_p$ for $\bvec{K}_{pp}$ evaluates to $0$ and is therefore trivial if the second argument is located on $\partial\Omega_u$, that on $\partial\Omega_u$ given by Eq.~\eqref{eq:BCNeumannP} requires some thought. For each point $\bvec{r}_b$ on $\partial\Omega_u$, one has
$\bvec{K}_{pp}(\bvec{r}_b + \epsilon\bvec{n},\bvec{r}') - \bvec{K}_{pp}(\bvec{r}_b,\bvec{r}') = \delta(\bvec{r}_b + \epsilon\bvec{n}-\bvec{r}') - \delta(\bvec{r}_b -\bvec{r}')$. Only the cases $\bvec{r}' = \bvec{r}_b$ and $\bvec{r}' = \bvec{r}_b + \epsilon\bvec{n}$ are relevant, for which one finds
\begin{align}
&\bvec{K}_{pp}(\bvec{r}_b + \epsilon\bvec{n},\bvec{r}_b + \epsilon\bvec{n}) - \bvec{K}_{pp}(\bvec{r}_b,\bvec{r}_b + \epsilon\bvec{n}) = \delta(0) \\
& \bvec{K}_{pp}(\bvec{r}_b + \epsilon\bvec{n},\bvec{r}_b) - \bvec{K}_{pp}(\bvec{r}_b,\bvec{r}_b) = -\delta(0).
\end{align}
Therefore, $\bvec{n}\cdot\nabla \bvec{K}_{pp}(\bvec{r}_b,\bvec{r}_b) = -\bvec{n}\cdot\nabla \bvec{K}_{pp}(\bvec{r}_b,\bvec{r}_b+ \epsilon\bvec{n})$ on $\partial\Omega_u$, which proves the above statement.

To conclude, if neither the value of pressure on $\partial\Omega_p$ nor that of its gradient on $\partial\Omega_u$ differ from the reference solution $p\st{ref}$, it is not necessary to compute the evolution due to pressure deviations.

\subsection{Strength of the propagators}
We define the strength of the velocity-velocity propagator by its zeroth moment
\begin{equation}
\bvec{k}_{\bvec{u}\bvec{u}}\spt{(0)}(\bvec{r},t) \equiv \int_{\Omega} d^3r' \bvec{K}_{\bvec{u}\bvec{u}}(\bvec{r},\bvec{r}',t)
\label{eq:zerothmoment}
\end{equation}
with analogous expressions for the other kernels. The propagator strengths of velocity deviations obey
\begin{align}
& \frac{\partial}{\partial t} \bvec{k}\spt{(0)}_{\bvec{u}\bvec{u}}
+ \bvec{k}\spt{(0)}_{\bvec{u}\bvec{u}}\cdot\nabla \bvec{u}\st{ref} + \bvec{u}\st{ref}\cdot\nabla \bvec{k}\spt{(0)}_{\bvec{u}\bvec{u}} = -\nabla \bvec{k}\spt{(0)}_{p\bvec{u}} + \nu\nabla^2\bvec{k}\spt{(0)}_{\bvec{u}\bvec{u}} \label{eq:integratedmomU}\\
&\text{Tr}\Big[\big(\nabla\bvec{k}\spt{(0)}_{\bvec{u}\bvec{u}}\big) \cdot \big(\nabla \bvec{u}\st{ref}\big)\Big] + \text{Tr}\Big[\big(\nabla \bvec{u}\st{ref}\big) \cdot \big(\nabla\bvec{k}\spt{(0)}_{\bvec{u}\bvec{u}}\big)\Big] = - \nabla^2 \bvec{k}\spt{(0)}_{p\bvec{u}}\label{eq:integratedpU}
\end{align}
with initial condition
\begin{equation}
\bvec{k}\spt{(0)}_{\bvec{u}\bvec{u}} (\bvec{r},t = 0) = \bvec{1}
\end{equation}
and boundary conditions
\begin{align}
\begin{rcases}
&\bvec{k}\spt{(0)}_{\bvec{u}\bvec{u}}(\bvec{r}\rightarrow \bvec{r}_b) = \bvec{1}\\
& \bvec{n}\cdot\nabla\bvec{k}\spt{(0)}_{p\bvec{u}}(\bvec{r}\rightarrow \bvec{r}_b) = \bvec{0},
\end{rcases} \,\, \text{for}\,\, \bvec{r}_b\in \partial\Omega_u
\end{align}
and
\begin{equation}
\begin{rcases}
&\bvec{n}\cdot\nabla\bvec{k}\spt{(0)}_{\bvec{u}\bvec{u}}(\bvec{r}\rightarrow \bvec{r}_b) = \bvec{0}\\
& \bvec{k}\spt{(0)}_{p\bvec{u}}(\bvec{r}\rightarrow \bvec{r}_b) = \bvec{0}
\end{rcases} \,\, \text{for}\,\, \bvec{r}_b\in \partial\Omega_p.
\end{equation}
The integrated propagation due to pressure deviations can even be written down explicitly. Because of 
$\bvec{k}\spt{(0)}_{\bvec{u}p} (\bvec{r},t = 0) = \bvec{0}$ and $\bvec{k}\spt{(0)}_{pp}(\bvec{r}_b \in \partial\Omega_p,t) = 1$, it follows in a straight-forward fashion that
\begin{align}
\bvec{k}\spt{(0)}_{\bvec{u}p}(\bvec{r},t) = \bvec{0}\label{eq:integratedmomP}\\
\bvec{k}\spt{(0)}_{pp}(\bvec{r},t) = 1\label{eq:integratedPP}
\end{align}
solve the volume-integrated versions of Eqs.~\eqref{eq:momp} and ~\eqref{eq:pp}.

Equations~\eqref{eq:integratedmomU} and \eqref{eq:integratedpU} are numerically rather tame with the initial condition being simply a constant instead of a sharp delta function, and Eqs.~\eqref{eq:integratedmomP} and \eqref{eq:integratedPP} even provide analytic solutions. Hence, they can be conveniently used to assess the accuracy of $\bvec{K}_{\bvec{u}\bvec{u}}$, $\bvec{K}_{\bvec{u}p}$, $\bvec{K}_{p\bvec{u}}$, and $\bvec{K}_{pp}$ obtained from the numerical solution of Eqs.~\eqref{eq:momU}, \eqref{eq:pU}, \eqref{eq:momp}, and \eqref{eq:pp}
by computing their volume integrals and comparing the results with $\bvec{k}\spt{(0)}_{\bvec{u}\bvec{u}}$, $\bvec{k}\spt{(0)}_{\bvec{u}p}$, $\bvec{k}\spt{(0)}_{p\bvec{u}}$, and $\bvec{k}\spt{(0)}_{pp}$.

\section{Case description and numerical setup}
\label{S:Setup}
We implemented the deviation-propagation methodology in the \myHighlight{open-source} framework of OpenFOAM~\cite{openfoam} \myHighlight{in a standard finite-volume fashion. For the investigated case, no specific handling of the Dirac delta functions in the initial and boundary conditions, which became Kronecker deltas with appropriate prefactors upon discretization, was necessary. However, for high Reynolds number flows, a dedicated treatment might be in order to control numerical diffusion}. The code and the case described below can be accessed in our repository~\myHighlight{\url{https://github.com/ParticulateFlow/pfmFOAM-public}}\myDelete{\cite{pfmfoampublic} upon request}.

In the present study, we applied our approach to the prototypical case of vortex shedding behind a cylinder of radius $r\st{cyl} = 1\,\meter$. For the sake of simplicity, we reduced to problem to two dimensions and studied a box with extensions $[-20\,\meter; 30\,\meter] \times [-20\,\meter; 20\,\meter]$, which was discretized into a mesh with 9200 cells. The Navier-Stokes equations were solved using the PISO algorithm with time steps of $\Delta t\st{CFD} = 0.01\,\second$, which led \myHighlight{to} Courant numbers well below 1.0 for the investigated inflow velocities in the range from $u\st{in} = 1.0\,\meter/\second$ to $u\st{in} = 1.2\,\meter/\second$. With a kinematic viscosity of $\nu = 0.01\,\meter^2/\second$, these values corresponded to Reynolds numbers Re $= 200$ to Re $= 240$.

By monitoring drag and lift coefficients, we determined that data sampling of the reference time series with $u\st{in} = 1.0\,\meter/\second$ could start after an equilibration time of $t\st{equil} = 180\,\second$. Due to the high degree of periodicity established after $t\st{equil}$, we recorded flow fields for approximately one shedding cycle of $t\st{sample} = 10\,\second$ with a sampling step $\Delta t\st{sample} = 0.01\,\second$ equal to the solution step size.

Next, the propagators were obtained. To this end, the reference time series was split into ten intervals with durations $t\st{pred} = 1\,\second$ each, and for each of these segments, the corresponding \myHighlight{$\bvec{K}_{\bvec{u}\bvec{u}}$, $\bvec{K}_{p\bvec{u}}$}\myDelete{propagators} were calculated \myHighlight{resulting in a database of ten reference states with an accompanying propagator each}. Equations~\eqref{eq:momU} and \eqref{eq:pU} were solved with time steps $\Delta t\st{prop} = 0.0025\,\second, 0.005\,\second, 0.01\,\second$, and $0.02\,\second$ to assess their convergence behavior.

\begin{figure*}[htbp]
\centering
    \includegraphics[width=8.4cm]{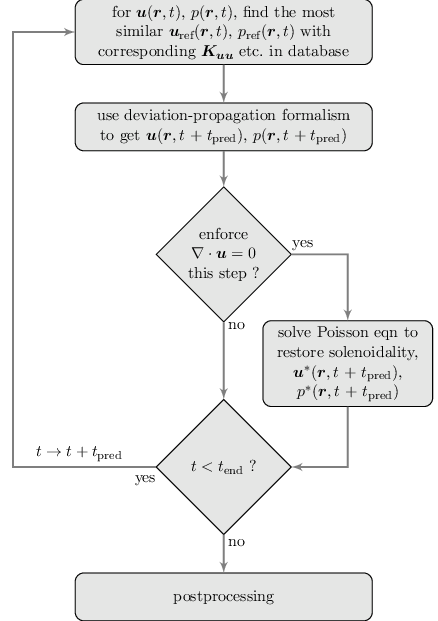}
	\caption{Flowchart of deviation-propagation predictions. Once the most similar reference state has been identified, the accompanying propagators evolve the deviations by $t\st{pred}$. If necessary, lack of solenoidality due to numerical errors can be fixed before proceeding to the next step or to postprocessing.} 
	\label{fig:flowchart}
\end{figure*}

The deviation-propagation approach was evaluated by comparing its predictions with full-CFD data for the following case variations:
\begin{enumerate}[label=\roman*]
  \item base case with uniform inlet velocity $u\st{in} = 1.0\,\meter/\second$
  \item uniform inlet velocity $u\st{in} = 1.1\,\meter/\second$
  \item uniform inlet velocity $u\st{in} = 1.2\,\meter/\second$
  \item same inflow as base case, rotating cylinder with $\omega\st{cyl} = \pi/4\,1/\second$
  \item same inflow as base case, rotating cylinder with $\omega\st{cyl} = \pi/2\,1/\second$
  \item same inflow as base case with superposed shear profile with $\partial u_x/\partial y = 0.2\,1/\second$
  \item same inflow as base case with superposed shear profile with $\partial u_x/\partial y = 0.4\,1/\second$ 
\end{enumerate}
For cases (vi) and (vii), the shear profile was limited to $-1\,\meter \leq y \leq 1\,\meter$, resulting in a slanted step function. For all of the above scenarios, predictions for a single step $t\st{pred}$ and for multiple ones were made. In the latter case, after each step, the most similar reference state within the set of ten configurations was determined by evaluating Eq.~\eqref{eq:errindicator} before the next prediction was made. The workflow is illustrated by Fig.~\ref{fig:flowchart}. Since the propagators were constructed to keep the flow divergence-free, \myHighlight{comparatively} few iterations were sufficient to solve the equation for the remaining pressure and restore full solenoidality.

We stress that for all case variations, the same database of reference states corresponding to the base setup was used. 

\section{Results}
\label{S:Results}
After a brief exploration of the general properties of the velocity-velocity propagator, results on its prediction capabilities for specific settings of the prediction step and cut-off value are presented. Afterwards, the influence of these parameters is highlighted.

\subsection{Propagators}
\begin{figure*}[htbp]
\centering
   \subfloat[\label{fig:magU}]{
   \includegraphics[width=8.4cm]{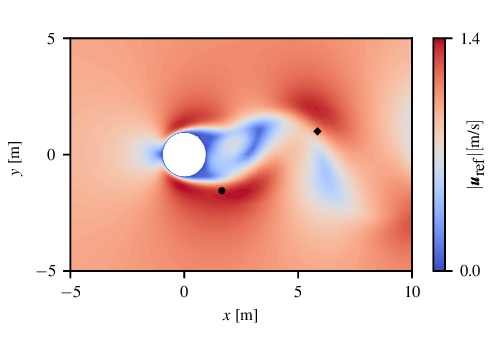}
   }\\
   \subfloat[\label{fig:Kuu_1}]{
   \includegraphics[width=8.4cm]{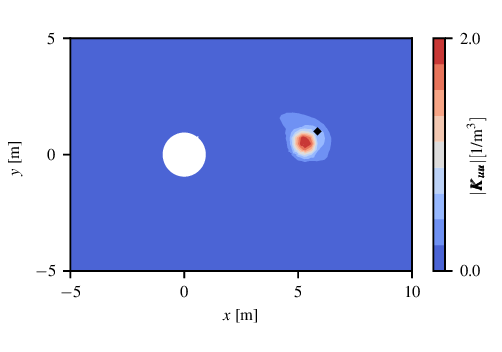}
   }\\
   \subfloat[\label{fig:Kuu_2}]{
   \includegraphics[width=8.4cm]{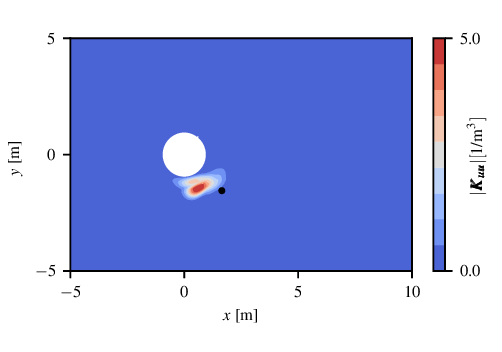}
   }
	\caption{Illustration of the flow and propagators. (a) shows a snapshot of the velocity magnitude with two selected reference points (indicated by a diamond and a circle) for which the corresponding propagator\myDelete{s} \myHighlight{norms according to Eq.~\eqref{eq:propnorm}} can be seen in (b) and (c).} 
	\label{fig:illustration}
\end{figure*}

\begin{figure*}[htbp]
\centering
   \subfloat[\label{fig:KminConv}]{
   \includegraphics[width=8.4cm]{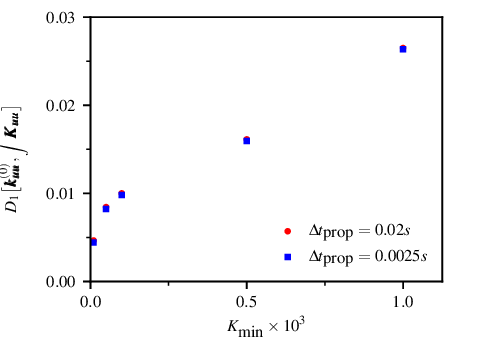}
   }
   \subfloat[\label{fig:DtConv}]{
   \includegraphics[width=8.4cm]{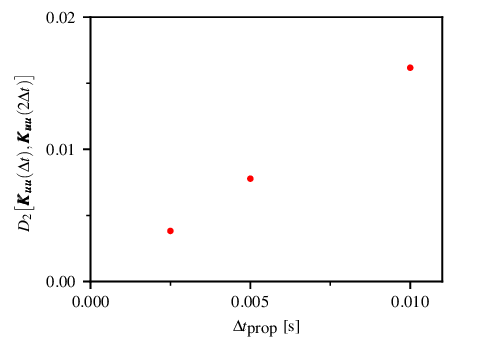}
   }
	\caption{Convergence of the velocity-velocity propagator. (a) The neglected strength of $\bvec{K}_{\bvec{u}\bvec{u}}$ decreased with decreasing $K\st{min}$. Below a critical value, the curve fell off rapidly. (b) The time step $\Delta t\st{prop}$ used for the calculation of the propagator with $K\st{min}=10^{-4}$ needed to be sufficiently small to reach an acceptable degree of convergence. For both (a) and (b), each data point was averaged over all reference states.} 
	\label{fig:covergence}
\end{figure*}

A first impression of the structure of deviation propagators is provided by Fig.~\ref{fig:illustration}, where the Frobenius norm
\begin{equation}
\big| \bvec{K}_{\bvec{u}\bvec{u}}\big|\myHighlight{(\bvec{r},\bvec{r}')} \equiv \sqrt{\text{Tr}\Big[\bvec{K}_{\bvec{u}\bvec{u}}^{T}\myHighlight{(\bvec{r},\bvec{r}')}\cdot\bvec{K}_{\bvec{u}\bvec{u}}\myHighlight{(\bvec{r},\bvec{r}')}\Big]}\label{eq:propnorm}
\end{equation}
is displayed for two selected points $\bvec{r}$. It can be seen how upstream points distributed around a small center of influence contributed to the propagator. With increasing distance from this center, the magnitude dropped off, which was a consequence of the diffusive term in the Navier-Stokes equations. For computational reasons, it was necessary to neglect any contributions $d^3r' {\big| \bvec{K}_{\bvec{u}\bvec{u}}\big|\myHighlight{(\bvec{r},\bvec{r}')} \leq K\st{min}}$ below a small, predefined threshold. Figure~\ref{fig:KminConv} demonstrates its impact in terms of the relative distance
\begin{align}
D_1 \big[ \bvec{k}\spt{(0)}_{\bvec{u}\bvec{u}}, \int\bvec{K}_{\bvec{u}\bvec{u}}\big] \equiv \frac{\int_{\Omega} d^3r \big|\bvec{k}\spt{(0)}_{\bvec{u}\bvec{u}}(\bvec{r}) - \int_{\Omega} d^3r' \bvec{K}_{\bvec{u}\bvec{u}}(\bvec{r},\bvec{r}') \big|}{\int_{\Omega} d^3r \big|\bvec{k}\spt{(0)}_{\bvec{u}\bvec{u}}(\bvec{r})\big|}
\end{align}
between the integrated propagator $\int_{\Omega} d^3r' \bvec{K}_{\bvec{u}\bvec{u}}(\bvec{r},\bvec{r}')$ and its strength $\bvec{k}\spt{(0)}_{\bvec{u}\bvec{u}}(\bvec{r})$ obtained from Eqs.~\eqref{eq:integratedmomU} and \eqref{eq:integratedpU}. 
Below a value of about $K\st{min}\spt{(crit)}\approx 10^{-4}$, $D_1$ decreased much more rapidly with decreasing $K\st{min}$ than above. This indicates that for larger values, the propagator was comparatively localized whereas for smaller ones, it was massively spread out.
We hypothesized that the former contributions were the physically relevant ones and discarded the latter ones as a compromise between accuracy and memory demands. Aside from the fact that they stemmed at least partly from numerical diffusion, their effect for flow field predictions would have been very small. Their already minor contribution of about $1\%$ to the overall strength of the propagator would have been further diminished because $\delta\bvec{u}$ could change its sign repeatedly over larger regions so that these parts of the convolution would have canceled out.

As shown by Fig.~\ref{fig:KminConv}, the integration time step $\Delta t\st{prop}$ had no impact on the strength of $\bvec{K}_{\bvec{u}\bvec{u}}$. However, it took a sufficiently small step size for the spatially resolved propagators to converge. This behavior can be observed in Fig.~\ref{fig:DtConv}, which shows the relative distance
\begin{align}
D_2\big[\bvec{K}_{\bvec{u}\bvec{u}}\spt{(1)},\bvec{K}_{\bvec{u}\bvec{u}}\spt{(2)}\big]\equiv
\frac{1}{V_{\Omega}}\int_{\Omega} d^3r \frac{\int_{\Omega} d^3r' \big|\bvec{K}_{\bvec{u}\bvec{u}}\spt{(1)}(\bvec{r},\bvec{r}')-\bvec{K}_{\bvec{u}\bvec{u}}\spt{(2)}(\bvec{r},\bvec{r}') \big|}{\sqrt{\int_{\Omega} d^3r' \big|\bvec{K}_{\bvec{u}\bvec{u}}\spt{(1)}(\bvec{r},\bvec{r}')\big|} \sqrt{\int_{\Omega} d^3r' \big|\bvec{K}_{\bvec{u}\bvec{u}}\spt{(2)}(\bvec{r},\bvec{r}')\big|}}
\end{align}
between propagators computed with a time step $\Delta t\st{prop}$ and with twice its value. In the present case, using the same time step as for the underlying CFD calculations led to an error below $1\%$. Given the approximate nature of the deviation-propagation approach, such a value seems to be acceptable, but a larger step size cannot be recommended.

\subsection{Predictions}
\begin{figure*}[htbp]
\centering
   \subfloat[\label{fig:singleStepPrediction}]{
   \includegraphics[width=8.4cm]{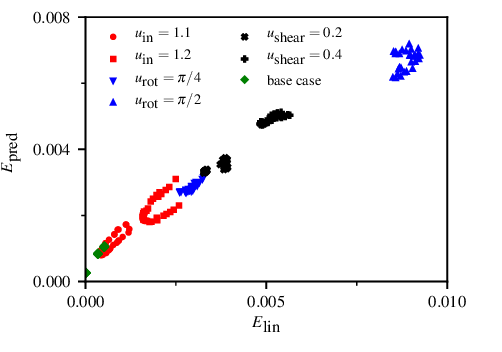}
   }
   \subfloat[\label{fig:multiStepPrediction}]{
   \includegraphics[width=8.4cm]{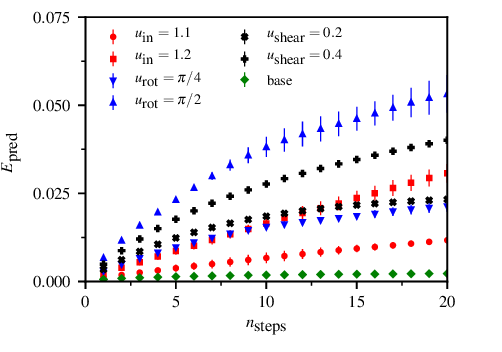}   
   }
	\caption{Single and multi step prediction errors. (a) After a single prediction step, errors for all cases were below $1\%$, depending on the magnitude of $E\st{lin}$. (b) The errors increased with several, consecutive steps. Error bars indicate a certain degree of scattering \myDelete{for some case variations depending on the}\myHighlight{obtained from 61 time series with different} initial state\myHighlight{s}.} 
	\label{fig:singleMultiStepPrediction}
\end{figure*}

To assess the quality of the deviation-propagation predictions, we first evaluated the relative error
\begin{equation}
E\st{pred} \equiv \frac{\int_{\Omega} d^3r \big|\bvec{u}\st{ex}(\bvec{r},t+t\st{pred}) - \bvec{u}\st{pred}(\bvec{r},t+t\st{pred})\big|}{\int_{\Omega} d^3r \big|\bvec{u}\st{ex}(\bvec{r},t+t\st{pred})\big|}\label{eq:relpredictionerror}
\end{equation}
between the exact solution $\bvec{u}\st{ex}$ and the prediction $\bvec{u}\st{pred}$ made by Eqs.~\eqref{eq:uansatz} after a single step\myDelete{ for a large number of times $t$ of the case variations described in Sec.~\ref{S:Setup}}. \myHighlight{For each of the case variations described in Sec.~\ref{S:Setup}, we used 61 flow fields drawn from a time series with sampling step $0.5\,\second$ and predicted the evolution of each of them.} Figure~\ref{fig:singleStepPrediction} shows a clear correlation with the error $E\st{lin}$ which was made in linearizing the Navier-Stokes equations. This demonstrates its capability to identify nearest neighbors.
The more similar a reference state from the prerecorded database could be found to a flow field for which a prediction should be made, the smaller the resulting error was. Notably, predictions with a step size of $t\st{pred} = 100\Delta t\st{CFD}$ were remarkably accurate. Even for the demanding case of a rotating cylinder, the relative error was below $1\%$.

It had to be expected that the error after multiple consecutive steps had accumulated. It can be seen in Fig.~\ref{fig:multiStepPrediction} that it grew the fastest for those cases where the single step predictions had been the most inaccurate. While for the base case (same boundary conditions but an initial condition not present among the reference states), hardly any deviation from the exact solution had established even after 20 steps corresponding to $20\,\second$ or approximately two shedding cycles, the case with the rotating cylinder showed a relative error of about $5\%$. However, the growth of the error had slowed down, and a similar behavior could be found for several of the other cases: a linear regime was followed by a slower increase. Even though the prediction accuracy was not extremely good anymore after this relatively long evolution time (compared with the underlying CFD time step), it seemed that 
long-term predictions with a much larger number of steps might be possible.

\begin{figure*}[htbp]
\centering
   \subfloat[\label{fig:meanFlow11}]{
   \includegraphics[width=8.4cm]{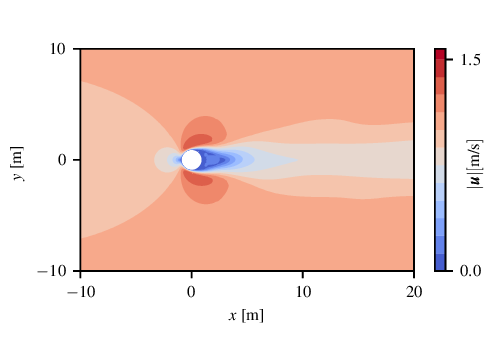}
   }
   \subfloat[\label{fig:meanFlow12}]{
   \includegraphics[width=8.4cm]{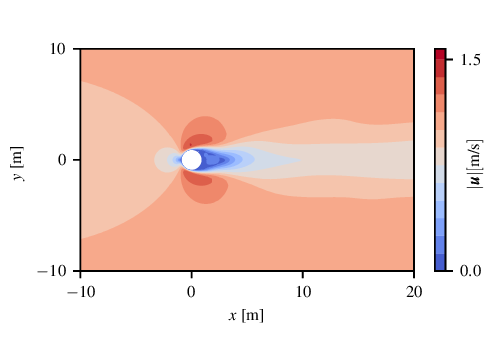}
   }\\
   \subfloat[\label{fig:meanFlow21}]{
   \includegraphics[width=8.4cm]{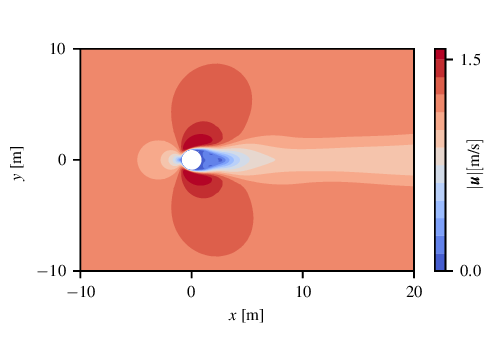}
   }
   \subfloat[\label{fig:meanFlow22}]{
   \includegraphics[width=8.4cm]{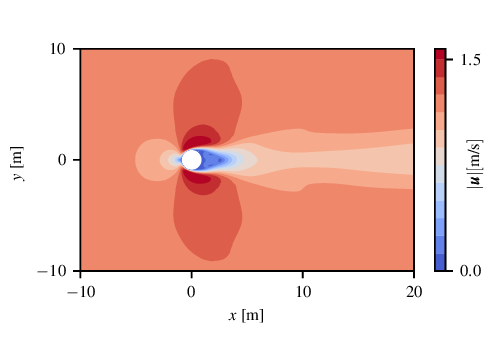}
   }
	\caption{Time-averaged flow fields. There were hardly any deviations for the base case (upper row) between full CFD simulations (left column) and deviation-propagation predictions (right column). A similar level of agreement was reached for the case of uniform inlet velocity $u\st{in}=1.2\,\meter/\second$ (lower row).} 
	\label{fig:meanFlow1}
\end{figure*}

\begin{figure*}[htbp]
\centering
   \subfloat[\label{fig:meanFlow31}]{
   \includegraphics[width=8.4cm]{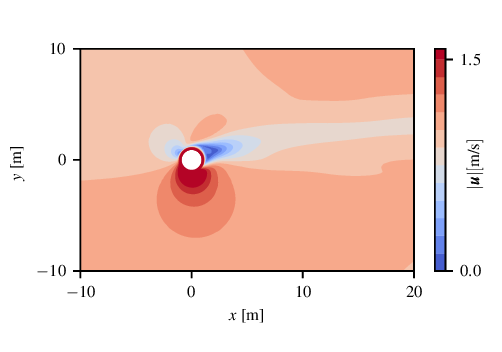}
   }
   \subfloat[\label{fig:meanFlow32}]{
   \includegraphics[width=8.4cm]{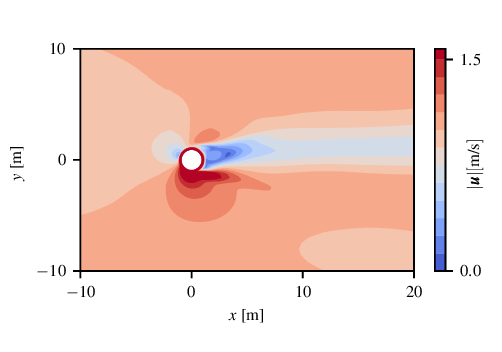} 
   }\\
   \subfloat[\label{fig:meanFlow41}]{
   \includegraphics[width=8.4cm]{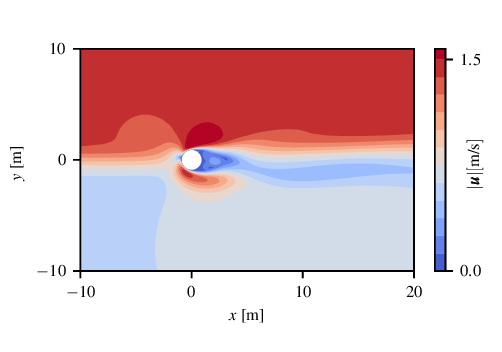}
   }
   \subfloat[\label{fig:meanFlow42}]{
   \includegraphics[width=8.4cm]{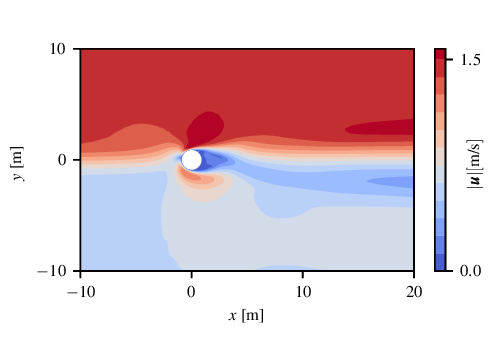}
   }
	\caption{Time-averaged flow fields under more demanding conditions. While local features of the exact solution (left column) differed from those predicted by the deviation-propagation formalism (right column), there were clear, qualitative similarities. Both a cylinder spinning with $u\st{rot}=\pi/2\,1/\second$ (upper row) and a shear inlet profile with $\partial u_x/\partial y = 0.4\,1/\second$ (lower row) led to a deflection of the vortex street that was comparable for the full CFD and the deviation-propagation simulation.} 
	\label{fig:meanFlow2}
\end{figure*}

This hope is indeed confirmed by Figs.~\ref{fig:meanFlow1} and \ref{fig:meanFlow2}, which show the time-averaged velocity fields for some of the investigated cases after $100\,\second$ corresponding to approximately ten shedding cycles. No deviation is visible at all for the base case in Figs.~\ref{fig:meanFlow11} and \ref{fig:meanFlow12}, and hardly can any differences can be found for the case with higher, uniform inlet velocity $u\st{in}=1.2\,\meter/\second$ in Figs.~\ref{fig:meanFlow21} and \ref{fig:meanFlow22}. A cylinder rotation of $u\st{rot}=\pi/2\,1/\second$ and a shear profile with $\partial u_x/\partial y = 0.4\,1/\second$ led to minor, quantitative differences that can be seen in Figs.~\ref{fig:meanFlow31} and \ref{fig:meanFlow32} and Figs.~\ref{fig:meanFlow41} and \ref{fig:meanFlow42}, respectively. However, the qualitative agreement was still very good considering that the database upon which predictions were made had the temporal average provided in Fig.~\ref{fig:meanFlow11}. In particular, it is noteworthy that the velocity fields close to the cylinder wall had very low values in the reference time series, whereas there were significantly larger velocities when the cylinder rotated or a shear profile was applied. Such conditions clearly challenged the validity of our linearization approach, which makes the prediction quality even more remarkable.

A more quantitative assessment of the prediction accuracy is provided by the relative errors along the lines of Eq.~\eqref{eq:relpredictionerror} for the time averages in Tab.~\ref{tab:errormeanflow}. Even the very demanding cases of high rotation rate and shear rate only caused relative errors of $3.5\%$ and $2.8\%$. The investigated uniform variations of the inlet velocity, on the other hand, led to deviations below $1\%$ and the base case had an utterly negligible error.

\begin{figure*}[htbp]
\centering
   \subfloat[\label{fig:aveU}]{
   \includegraphics[width=8.4cm]{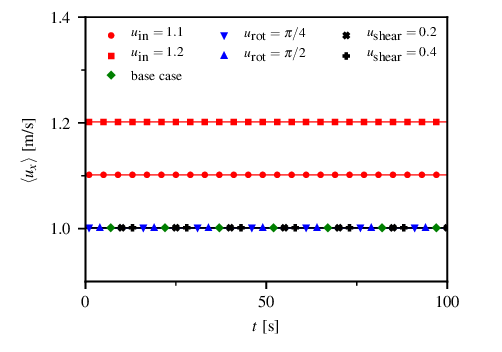}
   }
   \subfloat[\label{fig:aveU2}]{
   \includegraphics[width=8.4cm]{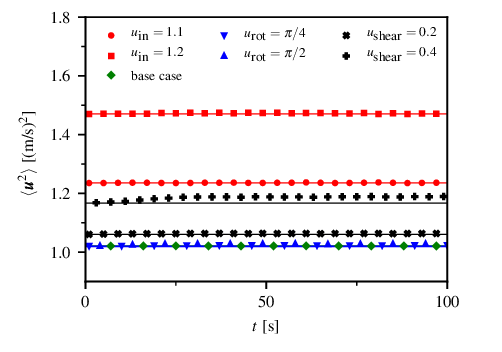}
   }
	\caption{Spatially averaged velocity and squared velocity. Straight lines resemble the results from full CFD calculations and symbols those from deviation-propagation predictions. No errors are visible regarding (a) the average velocity, and only a minor deviation for (b) the squared velocity of the shear-inlet profile can be seen.} 
	\label{fig:aveULong}
\end{figure*}

Besides the very good agreement of the long-term time averages of the predicted fields with the exact solution, it is also instructive to study the temporal evolution of global flow properties such as the volume-averaged velocity and squared velocity
\begin{align}
& \langle \bvec{u} \rangle(t) = \frac{1}{V_{\Omega}}\int_{\Omega} d^3r \bvec{u}(\bvec{r},t)\\
& \langle \bvec{u}^2 \rangle(t) = \frac{1}{V_{\Omega}}\int_{\Omega} d^3r \bvec{u}^2(\bvec{r},t).
\end{align}
In Fig.~\ref{fig:aveULong}, one observes an almost perfect conservation behavior with only one minor exception. The case with the higher shear rate had a slightly too high kinetic energy while all other cases fitted extremely well to the results obtained from a full CFD calculation.

\begin{table}[ht]
	\centering
	\begin{ruledtabular}
	\begin{tabular}{ll}
	case & rel.\ error\\
	\midrule
	base & 0.00065 \\
	$u\st{in} = 1.1\,\meter/\second$ & 0.0041  \\
    $u\st{in} = 1.2\,\meter/\second$ & 0.0083  \\
	$u\st{rot} = \pi/4\,1/\second$ & 0.014  \\
    $u\st{rot} = \pi/2\,1/\second$ & 0.035  \\
	$\partial u_x/\partial y = 0.2\,1/\second$ & 0.013  \\
    $\partial u_x/\partial y = 0.4\,1/\second$ & 0.028  \\
	\end{tabular}
	\end{ruledtabular}
	\caption{Relative error of time-averaged predictions.}
	\label{tab:errormeanflow}
\end{table}

\subsection{Influence of propagator cut-off and prediction step size}
\begin{figure*}[htbp]
\centering
   \subfloat[\label{fig:Xmin100}]{
   \includegraphics[width=8.4cm]{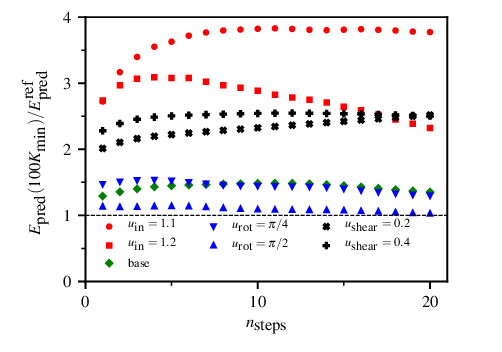}
   }
   \subfloat[\label{fig:Xmin10}]{
   \includegraphics[width=8.4cm]{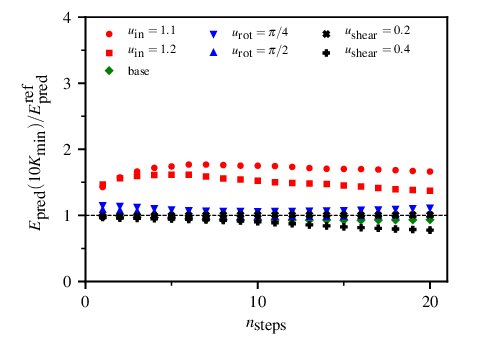}
   }
	\caption{Errors for different propagator cut-off values. Compared to predictions with a reference value of $K\st{min} = 10^{-4}$, those made with (a) $100K\st{min}$ were significantly larger for the cases with increased uniform and shear inflow. The base case and those with a rotating cylinder were less affected. (b) For $10K\st{min}$, errors were generally smaller, and only the configurations with increased uniform inflow showed significant deviations.
	} 
	\label{fig:Xminpred}
\end{figure*}

It is instructive to understand how the quality of the results presented in the previous section depends on the most important parameters of the propagators. As already discussed, it was necessary to impose a cut-off $K\st{min}$ below which contributions of the propagators were neglected. In the present case, we found $K\st{min} = 10^{-4}$ a reasonable choice. Figure~\ref{fig:Xminpred} demonstrates how a larger value impaired the prediction accuracy. It is not surprising that an extreme choice such as $K\st{min} = 10^{-2}$, where a significant portion of the propagator strength was missing, could cause much larger errors. Especially the cases of increased uniform and shear inflow velocity were strongly affected, while the others showed less severely increased errors. This observation may be explained by the fact that deviations for the former two types of flow would change their sign less rapidly as for a spinning cylinder. Hence, their contributions did not cancel out as easily. 

With $K\st{min} = 10^{-3}$, results were of similar quality as the reference with $K\st{min} = 10^{-4}$ except from those with increased uniform inlet velocity, where the effect mentioned above was the most pronounced. Since the same propagators were used for all case variants, it follows that a reasonable cut-off value depends on the target conditions to be simulated and not only on $\bvec{K}$ itself. For many cases, a generous choice of $K\st{min}$ can suffice while certain flow configurations require to retain contributions down to $K\st{min}\spt{(crit)}$. In order to reduce memory demands and computational costs, it is advisable to use as large a cut-off value as acceptable with regard to prediction accuracy.

\begin{figure*}[htbp]
\centering
   \includegraphics[width=8.4cm]{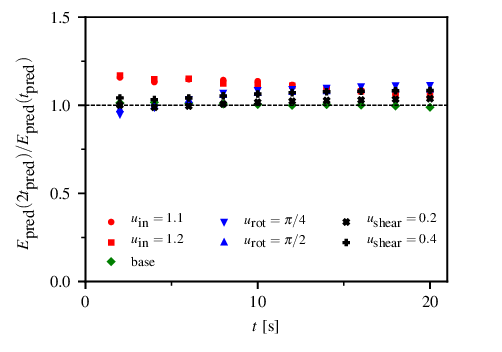}
	\caption{Errors for different prediction steps. Doubling the step size to $2t\st{pred}$ increased the prediction errors only moderately compared to those with $t\st{pred}$.} 
	\label{fig:tpred}
\end{figure*}

Besides the spatial range over which the convolution operations in Eqs.~\eqref{eq:uansatz} and \eqref{eq:pansatz} were carried out, the prediction step size influenced the quality of the obtained results. However, its effect was relatively weak. It can be seen in Fig.~\ref{fig:tpred} that making one prediction with $2t\st{pred}$ was only slightly more inaccurate than two with $t\st{pred}$. In the beginning, mainly the cases with increased uniform inlet velocity were affected. We believe a similar mechanism as discussed above was at work. With a larger prediction step size, the propagators became more extended, and the cut-off $K\st{min}$ led to slightly larger errors. After multiple steps, however, most case variants showed errors of similar magnitude. Besides the issue of decreasing propagator localization, the less frequent identification of the nearest neighbor state in the database might have had some impact. Due to the larger \myDelete{the }steps covered by the integral kernels, it took longer until deviations were computed from a different time series segment with a higher degree of similarity.

The above considerations demonstrate \myDelete{that it is inevitable to carry out}\myHighlight{the usefulness of} convergence and parameter variation studies for deviation-propagation simulations. 
Besides the issue of a reasonable compromise for $K\st{min}$, a sensible choice for $t\st{pred}$ needs to be made. Too large a value would lead to inaccurate predictions (together with massive memory demands for each propagator) whereas too small a value might not improve accuracy as much as desired, require a large number of reference states with accompanying propagators and impair the execution time.
\myHighlight{While they need not coincide, the temporal spacing between reference states (to be obtained with the aid of Eq.~\eqref{eq:lincriterion}) seems to be a plausible first guess for the propagator step size.}

\begin{figure*}[htbp]
\centering
   \includegraphics[width=8.4cm]{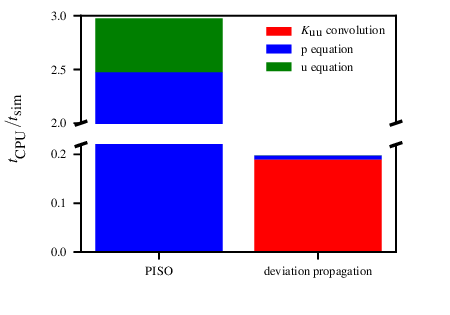}
	\caption{\myHighlight{CPU time per unit of simulated process time. For the PISO algorithm, the total numerical costs consisted predominantly of those for solving the pressure and the momentum equations. In the deviation-propagation simulations, most time was spent evaluating the convolution integrals and only a small fraction to restore solenoidality. The identification of nearest-neighbor states was negligible in the present case.}} 
	\label{fig:tCPU}
\end{figure*}

\myHighlight{
\subsection{Performance analysis}
\label{S:performance}
Although the present investigation has mainly demonstrated that a linearized propagator approach can approximate simple flows reasonable well, but did not attempt any optimization, a comparison with the runtime of the PISO algorithm is nevertheless instructive. Figure~\ref{fig:tCPU} reproduces the well-known fact that most of the computational effort of PISO-based simulations was due to the solution of the Poisson equation for pressure, while the momentum equation accounted for a significantly smaller fraction of CPU time. Remarkably, the numerical costs for deviation-propagation calculations were distributed in an opposite fashion. Almost all time was spent with the explicit step of computing the convolution integrals for large-step predictions whereas the implicit pressure correction required very little effort. Compared to its counterpart in the PISO algorithm, it had to be solved much less frequently due to the larger time step size, and it took fewer iterations per step because the propagators already accounted for incompressibility. The pressure correction mainly had to remove compressibility effects due to numerical inaccuracies.}

\myHighlight{
Overall, the deviation-propagation simulations with a time step of $t\st{pred} = 100\Delta t\st{CFD}$ ran about 15 times faster than the PISO calculations with standard settings. The former were about five times as fast as the simulated process, while the latter were three times slower. We did not attempt to explore the highest possible speed up at this point, but stress the very different contributions to runtime for both methods. Especially the fact that deviation-propagation calculations spent almost all CPU time in the explicit prediction step will allow for significant improvements in future work in terms of parallelization and a more efficient evaluation of the convolution integrals.
}

\begin{figure*}[htbp]
\centering
   \includegraphics[width=8.4cm]{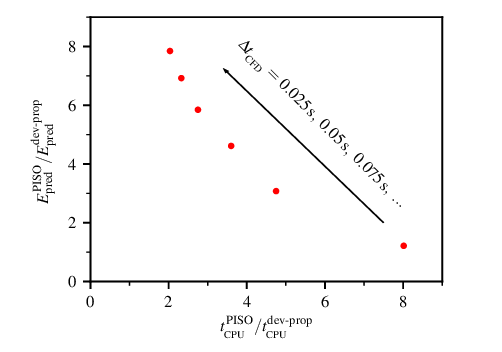}
	\caption{\myHighlight{CPU times and prediction errors for the PISO algorithm with different time step sizes. CPU times and errors are provided relative to those of deviation-propagation simulations for the base case. The reference velocity field towards which the errors were computed was obtained with $\Delta t\st{CFD} = 0.01\,\second$.}} 
	\label{fig:tPISO}
\end{figure*}

\myHighlight{
Even though a further, significant acceleration of the approach seems to be within reach, it is legitimate to ask if similar speed ups to those observed in this study might be achieved by simply decreasing the target accuracy of standard CFD algorithms. In Figure~\ref{fig:tPISO}, the deviation Eq.~\eqref{eq:relpredictionerror} of the long-term averaged velocity field obtained with the PISO algorithm using increasing time step sizes is shown together with the required CPU time for a fixed process duration. With sufficiently large steps, the runtime decreased towards that of deviation-propagation simulations. At the same time, the overall error increased strongly. At twice the CPU time, it was approximately eight times larger than that in the deviation-propagation calculations for the base case. This shows that a similar speed up cannot be bought easily for conditions present in the database. Of course, off-database simulations come with larger prediction errors where such a comparison would be more favorable for standard CFD algorithms, especially ones of higher order~\cite{Bijl2002,Forti2015,Loy2019}, with reduced accuracy requirements. Ultimately, the adequacy of the propagator approach will depend on how far away from the attractor in the database one wants to carry out simulations and how much one can further speed up the evaluation of the convolution integrals.}

\section{Conclusion and future work}
\label{S:Conclusion}
This work has introduced a novel, data-assisted and physics-informed combination of the method of analogues with a propagator approach for fluid mechanical problems.
Given an initial flow state, a similar reference configuration is identified in a prerecorded database. Since its temporal evolution is known and that of the deviation between the two can be obtained employing their propagator, predictions with very large steps are possible. Put differently, the non-linearity of the full Navier-Stokes equations is taken care of by the data-driven method of analogues while deviations from the reference states are governed by a linearized version of the Navier-Stokes equations, for which a large-step propagator can be computed.

For a number of case variations of vortex shedding, step sizes of up to 200 times larger than that used to generate the underlying data led to notably accurate results considering the magnitude of the prediction steps.
With regard to the practical applicability of the method, its long-term behavior was of specific interest. The obtained temporal averages after approximately ten shedding cycles were very close to full CFD results, and excellent conservation of global properties such as total momentum and kinetic energy was found. No divergence issues due to the potential accumulation of prediction errors arose.

The main purpose of our study was the introduction of a new simulation technique on a conceptual level. For this reason, we restricted ourselves to a small-scale test case. Due to the low number of cells,\myDelete{a detailed performance analysis and evaluation of the numerical costs would not have been representative and was therefore postponed}\myHighlight{ which is not necessarily representative for more demanding scenarios, we carried out only a rather basic performance analysis and evaluation of the numerical costs without going into intricate details. However, already these simple considerations demonstrated the explicit nature of the deviation-propagation formalism in contrast to many classical CFD algorithms and its potential for further improvements}.

\myDelete{However, b}\myHighlight{B}efore more demanding applications such as the study of temporal multi-scale problems of highly dynamic flows, their control or parameter variation/optimization involving a large number of simulations can be targeted, several steps will have to be taken. In particular, two issues are very pressing: (i) The computation of the propagators needs to be massively accelerated, and (ii) their representation has to become more efficient to reduce the tremendous memory demands. Both problems could be cured with \myHighlight{either an approximation of the propagators in terms of a few leading-order moments such as Eq.~\eqref{eq:zerothmoment} or} a clever choice of ansatz function with a few optimization parameters. As long as propagators are localized to a region of influence, they might be approximated, e.g., as (a sum of) Gaussians. Hence, for each point in the domain, one would only have to determine and store the corresponding small set of  characteristic parameters. If a more elaborate treatment turns out to be necessary, PINNs would offer greater flexibility to approximate the propagators.

Once the handling of the deviation-propagation approach has become more efficient, it will be interesting to apply it to more demanding conditions such as multi-scale and/or multiphase flows, where coupled propagator equations and/or higher-order terms beyond a first-order expansion need to be dealt with.






\bibliographystyle{elsarticle-num}

 
\end{document}